\documentclass[%
preprint,
 amsmath,amssymb,
 aps,
prc, showpacs,
showkeys
]{revtex4-2}

\usepackage{graphicx}
\usepackage{dcolumn}
\usepackage{bm}
\usepackage{epsfig}
\usepackage{subcaption}
\usepackage{booktabs}
\usepackage{hyperref}
\usepackage{amsmath}
\usepackage{float}
\usepackage{mathtools}
\usepackage{booktabs, multirow, array, makecell, caption}
\usepackage{rotating}
\usepackage[toc]{appendix}

\begin{document}
\title{Multi-step particle emission probabilities in superheavy nuclei at moderate excitation energies}

\author{A. Rahmatinejad}
\affiliation{Joint Institute for Nuclear Research, Dubna, 141980, Russia}
\author{T. M. Shneidman}
\altaffiliation[Also at ]{Kazan Federal University, Kazan 420008, Russia}
\affiliation{Joint Institute for Nuclear Research, Dubna, 141980, Russia}
\author{G. G. Adamian} \email{adamian@theor.jinr.ru}
\affiliation{Joint Institute for Nuclear Research, Dubna, 141980, Russia}
\author{N. V. Antonenko}
\affiliation{Joint Institute for Nuclear Research, Dubna, 141980, Russia and Tomsk Polytechnic University, 634050 Tomsk, Russia }
\author{P.~Jachimowicz}
 \affiliation{Institute of Physics,
University of Zielona G\'{o}ra, Szafrana 4a, 65516 Zielona G\'{o}ra, Poland}
\author{M.~Kowal}
\affiliation{National Centre for Nuclear Research, Pasteura 7, 02-093 Warsaw, Poland}

\begin{abstract}
The probabilities of $xn$-, $pxn$-, and $\alpha xn$-evaporation channels in excited superheavy nuclei were evaluated using the Monte Carlo method. The calculations utilized microscopically determined nuclear level densities and were compared with results obtained from the phenomenological Jackson formula. Effective temperatures derived from the microscopic approach were incorporated into the Jackson formula for different evaporation channels at low and moderate excitation energies. Additionally, an analytical formula was introduced to estimate the average kinetic energy of emitted particles in multi-step processes.

\end{abstract}

\pacs{21.10.Ma, 21.10.Pc, 24.60.Dr, 24.75.+i}

\keywords{Superheavy nuclei, microscopic-macroscopic model, fission.}

\maketitle


Superheavy nuclei (SHN) such as Cn, Nh, Fl, Mc, Lv, Ts, and Og, with charge numbers ranging from 112 to 118, have been successfully synthesized through complete fusion reactions induced by $^{48}$Ca projectiles and actinide targets. These reactions occur in the $xn$-evaporation channels, where $x$ represents the number of emitted neutrons. The literature sources \cite{Og1, Og1n,Og1nn, Og2U, Og2, SH1, SH2, SH3, SH4} provide detailed information on these reactions.

Most of these superheavy nuclei have been produced in the $3n$- and $4n$-evaporation channels, with excitation energies of the compound nucleus (CN) in the $E_{CN} = 30 - 40$ MeV. This achievement raises the question of expanding the table of nuclides even further. One possibility is the operation of the Factory of SHN at the Joint Institute for Nuclear Research (Dubna) \cite{Og2,Gulbekian19}, which could offer a new avenue for studying and producing new isotopes of superheavy nuclei at moderate excitation energies ($E_{CN} \ge 50$ MeV).

Nevertheless, the rate at which the survival probability of the CN and, consequently, the cross-section for evaporation residues decrease with the augmentation of beam energy remains a subject of ongoing inquiry \cite{Hong2021,Adamian2022}.

One of the main ingredients in calculating the survival probability
\begin{eqnarray}
W_s(U)=P(s,U)\prod_{i_s=1}^{x_{s}} \frac{\Gamma_{i_s}(U_{i_s})}{\Gamma_{t}(U_{i_s})}
\label{surv}
\end{eqnarray}
is the probability $P(s,U)$ of a particle evaporation channel
[a $xn$-evaporation channel ($s=xn$, $x_{s}=x$) or a channel with the charged particle ($p$ or $\alpha$) emission followed by $xn$-evaporation  ($s=p xn$ or $\alpha xn$, $x_{s}=x+1$)] at the excitation energy $U$ \cite{Vandenbosch1973}. In Eq. (\ref{surv}), the total width $\Gamma_{t}$ is the sum of widths of particle evaporation $\Gamma_{i_s}$ and fission $\Gamma_{f}$. The widths are calculated using the corresponding thresholds for decays and level densities \cite{Rahmatinejad2021,Zubov2002,Adamian2012, Nasirov2011, Wang2012,MichalK2019}.
The probability $P(xn,U)$, from which the positions of the excitation function  maxima crucially depend,  is calculated in the literature using the Jackson analytic formula \cite{Jackson}
\begin{eqnarray}
\label{eq14}
P(xn,U)=I(\Delta_{x},2x-3)-I(\Delta_{x+1},2x-1),\nonumber\\
I(\Delta_{x},2x-3)=1-\exp[-\Delta_{x}]\left[\sum_{i=0}^{2x-3}\frac{(\Delta_{x})^{i}}{i!}\right],
\end{eqnarray}
where
$I$ is Pearson's incomplete gamma function and $\Delta_{x}=\left(U-B_{xn}\right)/T$,  $B_{xn}=\sum_{i=1}^{x}B_{i}$, $T$ is the effective temperature, and $B_i$ is the neutron binding energy \cite{Jackson}.
The first incomplete gamma function gives the probability that the original CN will emit at least $x$ neutrons. The second function gives the probability that the residual nucleus has enough excitation energy to emit at least $x+1$ neutrons finally.
For the final residual nucleus after emission of $x$ neutrons, with an excitation energy  higher than the corresponding fission barrier height $B_f$ but insufficient for neutron evaporation, $B_{x+1}$ is replaced by $B_{f}$ in the calculation of $\Delta_{x+1}$ \cite{Vandenbosch1973}. In the case of exactly one neutron emission, Eq. \eqref{eq14}  is transformed into a simple formula
\begin{eqnarray}
\label{eq16}
   P(1n,U)=
\begin{dcases}
    1,& B_{1n}<U<B_{2n}\\
    e^{-\Delta_{2}}(1+\Delta_{2}),              & U\ge B_{2n},
\end{dcases}
\end{eqnarray}
because the probability of emitting at least $1$ neutron for $U>B_1$ equals one.

When calculating $P(s,U)$ for a channel in which a charged particle  (for example, an $\alpha$-particle or proton) evaporates, it is necessary to expand the expression for $P(s,U)$ taking into account the corresponding Coulomb barrier when calculating the corresponding value of $\Delta_{x}$ \cite{Zubov2003}.

Equation \eqref{eq14}  is derived by incorporating the three main assumptions: i) The neutron energy spectrum is
given by the Maxwellian distribution $\varepsilon_{i}e^{-\varepsilon_{i}/T}$, where $\varepsilon_{i}$
is the kinetic energy of the neutron. ii)
The neutron is necessarily evaporated if  the nucleus has enough excitation energy.
iii) The effective temperature $T$ is independent of excitation energy.
However, the last assumption becomes inaccurate for  $xn$-evaporation channels with $x\gg 1$.

The decrease in nuclear temperature in the evaporation process is taken into account in the calculations
by an effective temperature
\begin{eqnarray}
\label{eq16C}
T=T_{CN}/\sqrt{C},
\end{eqnarray}
where $C$ is a constant  and $T_{CN}$ is the temperature of the initial CN \cite{Zubov2003}.

The primary objective of this letter is to justify the utilization of the Jackson formula, incorporating the relation \eqref{eq16C}, while also determining the parameters $C$ for the $xn$-, $pxn$-, and $\alpha xn$-evaporation channels ($x$ = 1--9) of excited SHN within the framework of a microscopic approach. Another aim is to investigate the average kinetic energy of emitted particles at low and moderate excitation energies. This investigation offers valuable insights for estimating the survival probabilities and, consequently, the production cross sections of SHN across a wide range of excitation energies for the original CN. It can also improve the predictions for the cross sections.

The determination of $P(s,U)$ can be effectively achieved using the Monte Carlo (MC) method. In a multi-step process, MC sampling necessitates a probability distribution of particle kinetic energy to assess the feasibility of ejecting the next particle at each step. We define the probability density of emitting a particle with kinetic energy within a small interval centered around $\varepsilon_i$ as follows:
\begin{eqnarray}
\label{eq2}
p(\varepsilon_i,E_{i})=N \varepsilon_{i} \rho_{res}(E_i-B_i-\varepsilon_i) \delta\varepsilon_i,
\end{eqnarray}
where  $\rho_{res}(E_i-B_i-\varepsilon_i)$ is the nuclear level density (NLD) of the residue nucleus, and $N$ is the normalization constant.
Particle emissions depend strongly on the NLD of the corresponding final states.
Also, when a multi-step process becomes essential at high excitation energies, NLD is necessary to obtain additional information for calculating $P(s, U)$.  Several methods for calculating NLD are usually based on two approaches: combinatorial and thermodynamic \cite{Hilaire2001,Hilaire2006,Alhassid2000,Alhassid2007}. In this paper, the statistical formalism based on the superfluidity model \cite{origin1,Decowski1968,Bezbakh2014,Bezbakh2016,Rahmatinejad2020,Rahmatinejad2021,Rahmatinejad2022} is used to calculate the NLD of nuclei formed after the emission of a neutron, a proton, and an $\alpha$-particle. This method consistently considers pairing and shell effects when estimating the probabilities of $xn$- and $p xn$ or $\alpha xn$-evaporation channels.
The single-particle energies and  decay thresholds are obtained within the microscopic-macroscopic approach based on the deformed single-particle Woods-Saxon potential and Yukawa-plus-exponential macroscopic energy \cite{Jachimowicz2017_I,Jachimowicz2021}. As shown in Ref. \cite{Rahmatinejad2020}, the  model describes well the experimental NLD  for various Dy and Mo isotopes.
The CN excitation energy
\begin{eqnarray}
\label{eq4}
E_i=U-\sum_{j=1}^{i-1}\left(B_j+\varepsilon_j\right)
\end{eqnarray}
In the $i^{\textrm{th}}$ evaporation step, the probability of particle emission depends on the particle emission thresholds $B_{j}$ and the kinetic energies $\varepsilon_j$ carried away by the particles emitted in previous steps. We iterate the MC sampling process for consecutive decays until the nucleus depletes the required excitation energy for the subsequent step. During the calculation, we record and average the kinetic energies of the particles over the number of samples performed for each subsequent step in the $s$-evaporation channel.


\begin{figure}[h]
\centering
\includegraphics[width=0.5\textwidth] {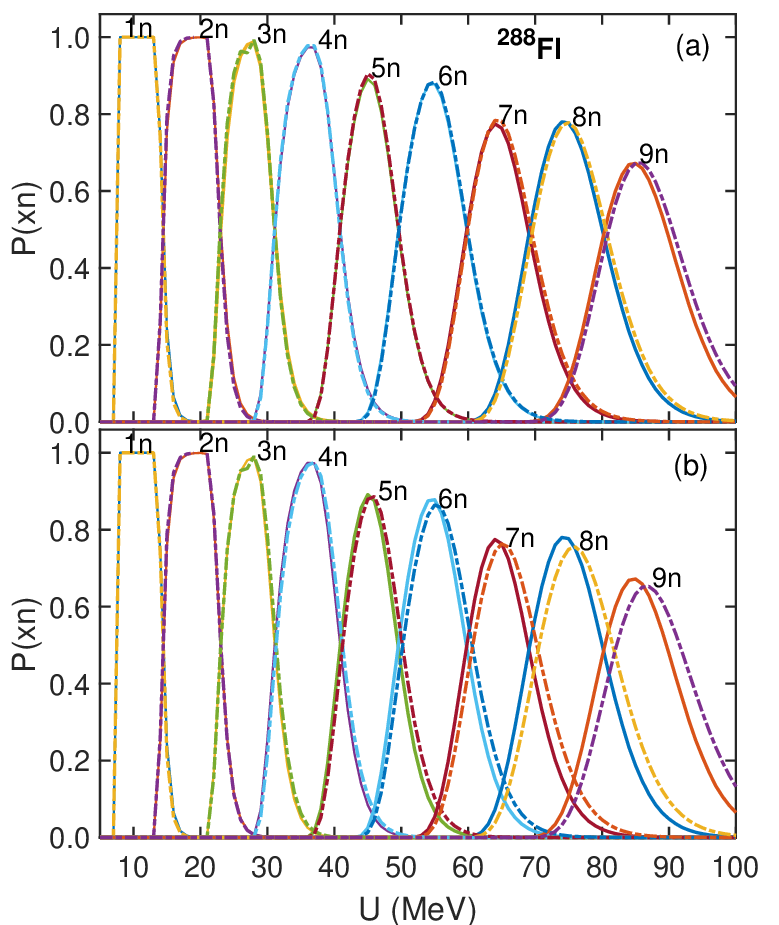}
\caption{Calculated probabilities of ($1n$--$9n$)-evaporation channels for the nucleus $^{288}$F as a function of excitation energy $U$. Solid lines are calculations within the microscopic approach; dashed
lines are calculations using the analytical formula \eqref{eq14}. The effective temperatures are taken as $T=T_{CN}/\sqrt{2}$ (a) and $T=\sqrt{U_{CN}/(2a_{CN})}$ (b).
The values of $T_{CN}$,  $U_{CN}$, and the level density parameter $a_{CN}$ for the  initial CN are   microscopically calculated.}
\label{Pxn-mic-MC-Gamma-2Fig-288-114}
\end{figure}

\begin{figure}[h]
\centering
\includegraphics[width=0.5\textwidth] {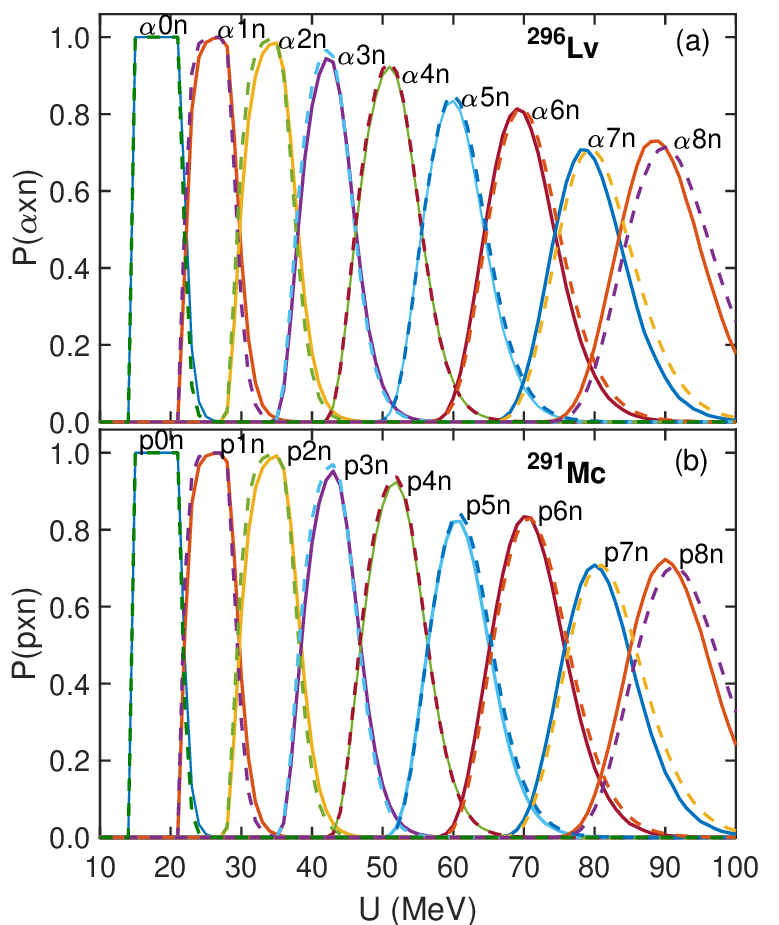}
\caption{Probability of  $\alpha xn$-evaporation channel ($x$=0--8) for $^{296}$Lv (a), and $pxn$-evaporation channel  ($x$=0--8) for $^{291}$Mc (b)   calculated within the microscopical approach (solid lines) and with the analytical formula   \eqref{eq14} modified for charged particle emissions (dashed lines) versus excitation energy $U$. The effective temperatures are taken as $T=T_{CN}/\sqrt{2.25}$ with the original CN temperatures $T_{CN}$.}
\label{PalphaPxn116-115}
\end{figure}

\begin{figure}[h]
\centering
\includegraphics[width=0.5\textwidth] {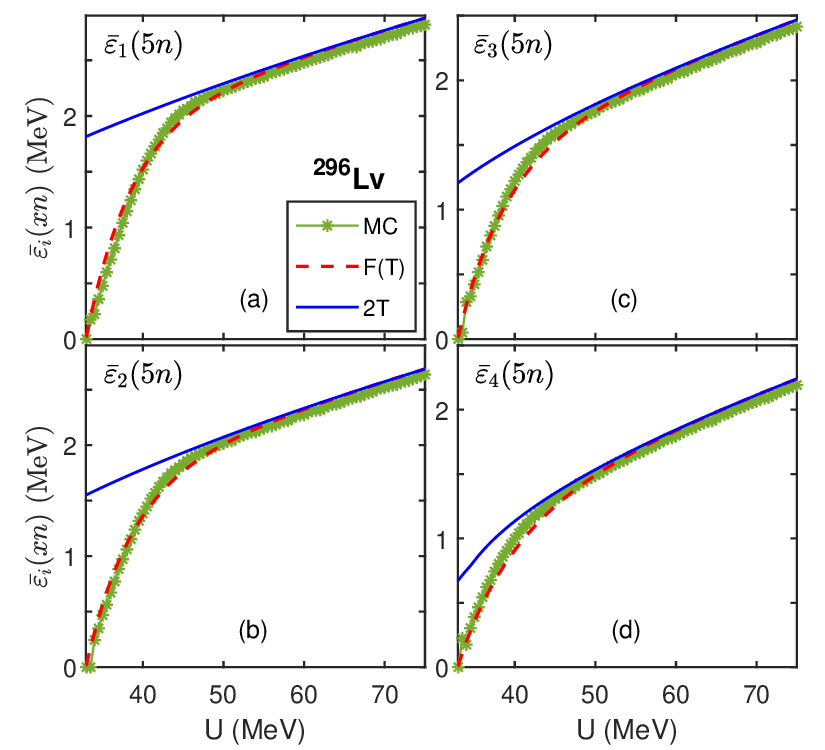}
\caption{For the $^{296}$Lv nucleus, the average kinetic energies carried away by neutrons at $i$=1--4 (a)-(d) intermediate evaporation steps in the $5n$-evaporation channel, calculated microscopically (green lines with asterisks) and with the formulas $2T_{i}$ (solid blue lines) and \eqref{eq25} (red dotted lines) depending on the excitation energy $U$.
}
\label{ei116}
\end{figure}

In Figs. ~\ref{Pxn-mic-MC-Gamma-2Fig-288-114} and \ref{PalphaPxn116-115}, the probabilities $P(s,U)$ of  $xn$-, $pxn$-, and $\alpha xn$-evaporation channels ($x$=1--9) in superheavy nuclei $^{288}$Fl, $^{291}$Mc and $^{296}$Lv at excitation energies of  (10 -- 100) MeV are calculated by MC method using Eq.~\eqref{eq2} and microscopic NLD. For comparison, $P(s,U)$ are calculated using the phenomenological Jackson formula \eqref{eq14}.
As can be seen for the $^{288}$Fl nucleus in $xn-$evaporation channels, the Jackson formula \eqref{eq14} with $C=2$ [see Eq. \eqref{eq16C}] gives results close to those of microscopic calculations.
The maximum values of $P(xn,U)$ for $x=1-5$ depend weakly on $C$.
However, due to the cumulative effect of temperature in Eq. \eqref{eq14} the results deviate slightly from microscopic calculations as $x$ increases. In Fig.~\ref{Pxn-mic-MC-Gamma-2Fig-288-114}~(a), the values of initial CN temperature $T_{CN}$ in Eq. \eqref{eq16C}   are taken from microscopic calculations.
One can also calculate the effective temperature  $T=\sqrt{U_{CN}/(Ca_{CN})}$ ($C=2$) within the Fermi-gas model,
where $U_{CN}$ and $a_{CN}$ are the excitation energy and the level density parameter of the parent CN, respectively.
For the $^{288}$Fl nucleus, this is shown in Fig.~\ref{Pxn-mic-MC-Gamma-2Fig-288-114}~(b).

In Fig.~\ref{PalphaPxn116-115}, we compare the Monte Carlo (MC) calculations of $P(\alpha xn,U)$ for $^{296}$Lv and $P(pxn,U)$ for $^{291}$Mc, where $x$ ranges from 0 to 8, with the corresponding values obtained using the modified analytical formula \eqref{eq14} for emitting charged particles.

For the $p xn$ or $\alpha xn$-evaporation channels, we find that the Jackson formula \eqref{eq14} with a slightly larger constant value of $C=2.25$ compared to that used for $xn$-evaporation channels provides a good reproduction of the values of $P(s,U)$ obtained from microscopic calculations. This can be attributed to the higher energy requirement for the emission of charged particles in the initial step, resulting in a significant decrease in temperature compared to neutron emission.

In each evaporation step $i$ with the corresponding nucleus temperature $T_i$, the average kinetic energy of the Maxwellian distribution is equal to twice the temperature: $\bar{\varepsilon}_{i}=2T_i$.
For the $^{296}$Lv nucleus, the excitation energy dependence of the average kinetic energy $\bar{\varepsilon}_{i}$ ($i$=1--4) of the neutrons in the $5n$-evaporation channel obtained using the MC method is presented in Fig.~\ref{ei116} together with $\bar{\varepsilon}_{i}=2T_{i}$.
From this figure, it can be concluded that moderate excitation energies support the assumption of the Maxwellian distribution. However, at low excitation energies, it becomes vital to consider the emission of neutrons with kinetic energies less than $2T_i$ in each evaporation step for the correct treatment of the channel's opening.
The expression

\begin{eqnarray}
\label{eq25}
\bar{\varepsilon}_{i}(xn,U)=2T_{i}\left(1-e^{-\frac{U-B_{xn}}{x}}\right)
\end{eqnarray}
for the average kinetic energy in the $xn$-evaporation
channel  leads to good agreement with the MC calculations over the entire energy range.
This expression is examined for $^{296}$Lv in Fig.~\ref{ei116} (red dashed lines).
The average kinetic energies for $p xn$ or $\alpha xn$-evaporation channels can be obtained similarly.


In conclusion, our study demonstrates that the analytical Jackson formula, with effective temperatures of $T=T_{CN}/\sqrt{2}$ and $T=T_{CN}/\sqrt{2.25}$, accurately reproduces the probabilities of $xn$,  $p xn$, and $\alpha xn$-evaporation channels in moderately excited SHN, as calculated using the microscopic model.

Furthermore, we derived an analytical formula \eqref{eq25} for the average kinetic energies of emitted particles. This expression provides improved results compared to those obtained using the Maxwellian distribution, particularly at low excitation energies in intermediate steps of the evaporation chain, especially for many-particle emissions.

Moreover, our findings indicate that the accuracy of the calculated probabilities of $xn$, $p xn$, and $\alpha xn$-evaporation channels, as well as the average kinetic energies of emitted particles, can be enhanced by using the temperature $T_{CN}$ of the initial CN calculated within the Fermi-gas model with an energy-dependent level density parameter.

Overall, our results contribute to a better understanding of particle emission processes in SHN and provide valuable insights for more accurate predictions of evaporation channels and  properties of emitted particle that play a crucial role in predicting the survival of newly synthesized SHN.

{\bf Acknowledgements:}

M.K. was co-financed by the National Science Centre under Contract No. UMO-2013/08/M/ST2/00257 (LEA COPIGAL).
T.M.S., G.G.A.,  and N.V.A. were supported by Ministry of Science and Higher Education of the Russian Federation (Moscow, Contract No. 075-10-2020-117).

\end{document}